# Effects of discretization on viscosity in the necklace model

D.A. Mártin[1] and C.M. Aldao[2]

[1]Physics Department, School of Exact and Natural Sciences
Universidad Nacional de Mar del Plata
Deán Funes 3350, B7602AYL Mar del Plata, Argentina

[2]Institute of Materials Science and Technology (INTEMA)
Universidad Nacional de Mar del Plata-CONICET
Juan B. Justo 4302, B7608FDQ Mar del Plata, Argentina

## ABSTRACT

The consequences of discretization on the resulting viscosity for the necklace model are analyzed to find out the mechanisms by which this model leads to results different from what could be anticipated. In order to accomplish this, various models of increasing complexity are proposed. First, the effects of the jump distance and temporal discretizations are studied. Then, the lack of proportionality between the length and the mass of the chains and the fact that chains do not perform a random walk are analyzed. Knowing of these influences on the viscosity, we finally address the effect of length fluctuations. It is shown that fluctuations can have unexpected effects that lead to larger or smaller values of viscosity. The simplicity of the studied models allows us to quantify the effects of discretization and fluctuations, effects that can be present in other models of similar or larger complexity.

# Introduction

In concentrated solutions and melts, long polymer chains form an entangled threadlike group of objects. The fact that a chain of this "spaghetti plate" cannot intersect itself or other chains is responsible for a type of dynamics whose details continue being of interest after decades. Since every chain is constrained to move in the direction of the chain backbone, the motion resembles that of a snake, hence the name reptation. (The reptation concept has been a very successful attempt to describe the dynamics of entangled linear polymers that includes the topological effect of entanglement constraint [1, 2].) The chain movement is then restricted to a curvilinear tubelike region. This "tube" represents the skeleton of the chain under consideration that diffuses through slip links or entanglement points with other chains. Thus, the many chain problem is treated as a single chain in a mean field due to the rest of the chains. The original model predicts a tube diffusivity $D_{tube}$ that scales with the molecular weight $M$ as $1/M$, a diffusivity that scales as $1/M^2$, and a zero-shear-rate viscosity that scales as $\eta_0 \sim M^{\beta}$, with $\beta=3$.

The behavior of highly entangled polymeric liquids has been experimentally examined for years. It has been found that the zero-shear-rate viscosity exhibits a dependence $\eta_0 \sim M^{3.4}$. After many years, the discrepancy with the viscosity scaling predicted in the reptation model remains a controversial subject [3]. Two main modifications have been proposed to be responsible for these differences. First, the scaling observed in experiments could be attributed to a chain spring like motion, or contour length fluctuation (CLF), that speeds up relaxation in such a way that the viscosity including fluctuations is smaller than the viscosity when fluctuations are not incorporated [4, 5]. Second, all the chains are mobile and then entanglements are not permanent on the time scale of reptation, which implies a tube motion or constrain release (CR) [6-12]. In experiments, tube motions can be mostly suppressed by studying the dynamics of a small fraction of short entangled probe chains (tracer diffusion) in a high-$M$ matrix of low diffusivity [3, 13, 14]. In these studies, a value of $\beta>3$ was also determined but smaller than in self-diffusion experiments (in which the probe chain has the same molecular weight as the matrix in which it diffuses). Also, the tracer diffusivity behaves as $1/M^2$ as predicted in the original reptation model.

Chains can be represented with very different degree of detail. However, the tube idea suggests that coarse graining at the level of the spacing between entanglements, keeping chain connectivity and chain uncrossability, would retain the basic information for rheological investigations [15]. Uncrossability can be imposed by constraining inner particles to move along the tube that represents the skeleton of the chain and end particles are free to move in the 3D embedding space. In 1987, Rubinstein published a discretized numerical model of the reptation model, known as repton model [16] that presents an exponent $β>3$. Years later, in 2002, Guidoni and co-workers introduced the necklace model [17, 18], presenting a greater flexibility in choosing the jumping frequencies. Both models are one-dimensional discretizations that mimic the diffusion of a chain including CLF but not CR.

The necklace model presents a hard-core constraint in which beads cannot occupy the same site. Conversely, in the repton model any number of beads can be at the same site. Since a bead represents an entanglement strand, interpenetration should be possible but limited due to intrachain excluded volume. In this sense, the necklace and the repton models are two extreme limits. (Reality probably lies between them.) Although this type of discretization of the original reptation model is quite mature, the resulting viscosity shows a non-trivial behavior that has not been explained yet.

It is not the goal of the present work to reproduce experimental results but to determine the influence on viscosity of the rules used in a discretized version of reptation. Other mechanisms besides CLF are known to affect viscosity (such as the mentioned CR, correlations with the motion of other chains, or fluctuations in the tube diameter) but they will not be considered. In fact, our goal is to determine what is responsible, and in what extent, for the viscosity behavior in the necklace model. To accomplish this, we resort to a set of numerical models with increasing complexities. Thus, we will determine how the model characteristics affect the resulting mass dependence of the viscosity. The involved mechanisms could have their counterparts in other models and also in experiments. Finally, the resulting viscosity is compared with that of the repton model.

**Basic considerations**

The scaling found in the original reptation model can be readily derived using the Einstein relation [1]. If all beads have the same behavior, the frictional force is proportional to the number of beads in the chain, $N$. Then, the mobility $\mu$ must be equal to $\mu_1/N$, where $\mu_1$, independent of $N$, is the mobility of a single bead. The Einstein relation states that mobility and diffusion are proportional, therefore the one-dimensional diffusion coefficient or tube diffusivity, $D_{tube}$, must be equal to $D_1/N$, where $D_1$ is the diffusion coefficient of a single bead; i.e., $D_{tube} \propto 1/N$. To escape from the original tube, the chain must progress by tube diffusion a distance $L$, which is the length of the chain. Assuming that $L$ is proportional to $N$, the time needed for that is

$$\tau \propto \frac{L^2}{D_{tube}} = \frac{NL^2}{D_1} \propto N^3. \tag{1}$$

Thus, the time needed for the chain to abandon the initial tube, and then the viscosity $\eta$, is proportional to $N^3$. Note that contour length fluctuations and constrain release or tube reorganization are not included in the original reptation model.

The relaxation process can be studied by following the rates at which the chain vacates initially occupied sites [16, 19]. $l(t)$ will be the part of the original tube that has not met the chain ends during the time intervals between 0 and $t$. The zero-shear-rate viscosity can be calculated by integrating the stress, which is proportional to the fraction of the chain that remains within the original tube, *i.e.*

$$\eta_0 = \frac{1}{\langle L \rangle} \int_0^\infty \langle l(t) \rangle dt, \tag{2}$$

where $L$ is the chain length and the brackets denote the ensemble average.

## The necklace model

The model consists of a one-dimensional chain with $N$ beads or particles. Hardcore interactions are incorporated by considering that beads can hop to a nearest site only if this site is empty. Particles can hop to right or left but no more than a site can be empty between two particles; thus, chain integrity is preserved. The model only distinguishes end particles from middle particles. Indeed, the probability of hopping is the same for all the particles except for those at the ends. A middle particle jumps with a probability per unit time $p_c$ while end particles are allowed to jump with probabilities per unit time $p_a$ and $p_b$ when jumping stretches or compresses the chain, respectively. Hence, $p_a$, $p_b$, and $p_c$ are the free parameters in the necklace model. In Fig. 1, a chain of five beads is sketched showing possible bead jumping. In the following, we will assume that the distance between adjacent sites of the lattice and the unit time are both equal to 1.

According to the above rules, in the necklace model the distance between beads can be 1 or 2. If the distance between two consecutive beads is 2, there is an empty site between them, what is named a hole. The probability of having a hole between two particles can be easily calculated as shown in Ref. 18

$$P_h = \frac{p_a}{p_a + p_b}. \tag{3}$$

Note that, in equilibrium, $P_h$ must be the same along the whole chain. Then, the average number of holes in a chain is $P_h(N-1)$ since there are $(N-1)$ positions available for holes. Then, the average length of a chain is given by

$$\langle L \rangle = N + P_h(N-1), \tag{4}$$

Recently, the exact analytical expression for the diffusion coefficient for the one-dimensional necklace model was obtained [20, 21], (see also [22] where related results were obtained). It is found that the diffusion coefficient is strictly proportional to $1/N$ if $p_a+p_b=p_c$. Under this condition, all particles behave similarly and then the diffusivity presents the scaling originally predicted for reptation, $D_{tube} \propto 1/N$, specifically

$$D_N = \frac{p_a p_b}{p_c} \frac{1}{N}. \tag{5}$$

Note that the necklace model has in principle three parameters: $p_a$, $p_b$, and $p_c$. However, what is relevant in our studies is the ratio among them and thus we chose $p_c=1$. This reduces the number of free parameters to two. Finally, since we will only study the cases for which $p_a+p_b=p_c$, the model has only a single parameter, say $p_a=P_h$.

In principle, the basic difference between the original model of reptation, predicting cube molecular weight dependence of viscosity, and the necklace models is that in the original model of reptation a tube of constant length is assumed and then reptation becomes a random walk of a fixed length object. Conversely, in the necklace model the length of the tube fluctuates. The consequences of tube length fluctuations are discussed in Doi and Edwards book [2], but we will see that the results obtained with the necklace model are quantitative and qualitative different from Doi and Edward's treatment. However, there is no contradiction because we are dealing with a model that presents some details in its dynamics not previously studied.

In what follows, we will introduce simplified models to determine the consequences on the viscosity of the rules adopted in the necklace model.

## Simplified models

### Rigid chain model

This model is a direct discretization of the original reptation model. It consists of a chain of constant length $L=N$ moving as a whole and performing a random walk. Having in mind the necklace model, every time a particle moves the center of mass changes its position $\Delta x=1/N$. Accordingly, in the rigid chain model (RCM), we move the chain so that its center of mass moves $1/N$ without any chain deformation.

Within an interval $\Delta t$, the probability for a particle jump is, in average, $P=p_0 N \Delta t$, where $p_0$ is a constant. This is strictly correct in the limit $\Delta t \rightarrow 0$, when it is very unlikely that two jumps occur in the same interval $\Delta t$. In this model the chain has always the same probability of jumping to the right or to the left (jumps are not correlated) and the chain does not present length fluctuations. Note that the diffusivity scales as $1/N$ as in the original reptation theory. Indeed, the jump frequency of the chain is given by $k=P/\Delta t=p_0 N$ with a jump distance $\Delta x$. Since the chain performs a random walk, the diffusivity is given by $D=k(\Delta x)^2$, and then $D\sim 1/N$.

In Fig. 2 we show the resulting viscosity as a function of $N$ for a jump distance $\Delta x=1/N$ determined as described above, see Eq. (2). The straight line is proportional to $N^3$ showing that for large values of $N$ the viscosity behaves as expected. In order to stress how far from 3 the exponent $\beta$ is, in the inset of Fig. 2 we plot $\eta_0/N^3$ as a function of the chain length $N$. For large values of $N$ ($N>10$) the curve becomes horizontal indicating that $\beta\rightarrow 3$ in accord with the original reptation model. However, for small $N$, the viscosity has a higher value implying that $\beta$ is smaller than 3. (From now on, due to its sensitivity, we will present this type of plots.)

In Fig 3, we present similar results to those of Fig. 2 for different jump distance $\Delta x$ keeping the diffusivity constant. As we reduce the jump distance, $\eta_0/N^3$ converges faster to the same final value. Then, we conclude that the jump distance discretization is responsible for the anomalous viscosity observed for small values of $N$. This effect is not difficult to explain. To compare viscosities, chains must have the same diffusivity. Hence, as the jump distance is reduced, the jumping frequency must be increased so that $D=k(\Delta x)^2$ presents the same value for a given value of $N$. Note that, for example, a reduction in a factor of 2 of $\Delta x$ implies a value of $k$ four times larger. As a consequence, even though the diffusivity is kept constant, a chain with a smaller jump distance has the chance of exploring a wider region and then the integrand in Eq. (9) reduces faster. In other words, the mass center can have a similar displacement but a smaller spatial discretization implies more fluctuations that facilitate the escape from the original tube.

Since chains in the necklace model are made of beads that jump fixed lengths, we cannot reduce the jump distance and then we cannot elude this discretization effect. This effect will raise viscosity for small values of $N$ in this type of discretized versions of reptation.

We also checked the effects of temporal discretization and we found that it is of little importance. Indeed, we numerically calculated the viscosity using the largest step time and with a step time four times shorter (keeping the diffusivity constant). Also, instead of attempting to move the chain at regular periods, we moved it after a random period of time according to the Kinetic Monte Carlo method (for a review of this method, see [24, 25]). Differences in results were smaller than 5% for $N=2$ and $N=3$ and were not noticeable for higher values.

**Equivalent length model**

In the equivalent length model (ELM) we take a rigid chain that performs a random walk but its length is given by Eq. (4) that corresponds to the necklace model. Note that the chain length $L$ is not exactly proportional to the number of particles $N$; this can originate a non-negligible effect.

We have seen that the time needed to escape from the original tube ($\tau$), and then the viscosity, is proportional to $L^2/D_{tube}$, Eq. (1). Assuming that the tube diffusivity is proportional to $1/N$ and the length is proportional to $N$, the viscosity becomes proportional to $N^3$. However, when $L$ is not proportional to $N$, this is not longer valid. Using Eq. (4), we can estimate the influence on the viscosity due to the lack of proportionality between $L$ and $N$ in the necklace model for different values of $P_h$:

$$\left.\frac{L^2 N}{N^3}\right|_{P_h=0.9} \propto 1 - \frac{0.947}{N} + \frac{0.224}{N^2}, \tag{6}$$

$$\left.\frac{L^2 N}{N^3}\right|_{P_h=0.1} \propto 1 - \frac{0.182}{N} + \frac{0.0083}{N^2}. \tag{7}$$

In both cases, for $P_h$=0.9 and 0.1, the second term dominates and the resulting slope is always positive. Monte Carlo results of Fig. 4(a) show the effects of introducing the nonlinearity between $L$ and $N$ in the necklace model. The lines indicate the effects expected using Eqs. (6) and (7). Note that Monte Carlo results show the combined effect of discretization and nonlinearity between $L$ and $N$. For $P_h$=0.1, the effect of discretization on $\eta_0/N^3$ is mostly compensated in the range $20 \leq N \leq 100$ so that the slope is almost null. For $P_h$=0.9, the slope is positive as the equivalent length effect is stronger than that due to the discretization in jump distance. Indeed, the slope in the range $20 \leq N \leq 100$ is $\approx 0.03$.

In order to separate the effects due to the lack of proportionality between $L$ and $N$ from those due to the discretization in jump distance, in Fig. 5 we present results with smaller spatial discretizations for the ELM using the lengths corresponding to the necklace model ($P_h$=0.9). As seen in Fig. 3, discretization effects are almost suppressed for $\Delta x=1/10N$ and this directly reflects on the Monte Carlo results of Fig. 5. Therefore, we conclude that the difference between the calculated values and what is predicted with Eq. (6) is due to jump distance discretization.

**Hybrid model**

In general, in the necklace model (unless $p_a+p_b=p_c$), for a specific configuration, the probabilities of jumping in one or other direction are not the same. Moreover, the action of a moving particle at each step depends on the previous moves of the chain. It can be shown that a jump is more likely to occur in the opposite direction than the previous one, and the shorter the chain the stronger this correlation is [18]. As a consequence, the chain needs more time to abandon the original tube and then the viscosity is larger.

In the necklace model with $p_a+p_b=p_c$, end beads have the same probability to jump as the rest of the beads but chains do not perform a random walk: jump frequencies depend on the chain configuration, so, in more stable configurations, jumps are less likely to take place.

Diffusivity differences can be easily adjusted, but ratios between jump frequencies in different configurations are inherent to the model.

To check the influence on viscosity, we tested a model (the hybrid model, HM), which is similar to the ELM but moves as dictated by the center of mass in the necklace model. We compared the HM and ELM results and found that the viscosity is only significantly affected for very short chains, $N \leq 3$.

**Viscosity in the necklace model**

In Fig. 6, the numerically calculated values of the viscosity in the necklace model are plotted; case (a) (full circles) corresponds to $P_h$=0.9 and case (b) (full squares) to $P_h$=0.1. For small values of $N$, the large values of viscosity can be attributed to the jump distance discretization, effect that is partially compensated by the lack of proportionality between $L$ and $N$.

For $P_h$=0.9, in the range 20<$N$<100, $\beta$ adopts an effective value of 3.1 that we feel compelled to attribute to fluctuations. Indeed, in Fig. 6 we also present results for the hybrid model; the viscosity, which includes discretization and length effects but not fluctuations, is always larger than that for the flexible chain. This is the expected effect, as fluctuations imply an extending and compressing chain while diffuses and then an acceleration of the stress relaxation. As fluctuations are relatively less important for long chains, eventually, for large enough values of $N$, their effects vanish and then the viscosity for both models converge. In the range 20<$N$<100, the hybrid model presents a value for $\beta$ very close to 3; this is similar to what is observed in the ELM and then we attribute the small observed slope to the lack of proportionality between $L$ and $N$.

Proportionality between $L$ and $N$ can be easily achieved in a model of rigid chains by adopting a chain length proportional to $N$. In some chains consisting of beads, this can also be accomplished. Within the necklace model, we could enlarge $P_h/2$ the end beads, so that $L'=N+P_h(N-1)+P_h=N(1+P_h)$. We simulated this "long ends necklace model" and found the expected results: an increase in viscosity for small $N$ as predicted with Eqs. 6 and 7.

The next results are very surprising as reported in Ref. 23. Chains for which $P_h$=0.1 present the same diffusivity and amplitude of fluctuations than those corresponding to $P_h$=0.9. Conversely, for the same value of $N$, chains in case (b) are shorter because the number of holes is smaller. In the range $20 \leq N \leq 100$, Fig. 6 shows that $\beta$=2.96, smaller than 3. We also studied the consequences of eliminating length fluctuations using the hybrid model. The resulting exponent $\beta$ is practically 3 for $20 \leq N \leq 100$. These results imply that fluctuations can reduce $\beta$, which is a very unexpected effect.

We have recently explained these results noting that for $P_h \neq 0.5$, the chain length distribution is asymmetric with respect to the average length and the chain does not spend the same time in every configuration [23]. It can be observed that for chains with a small number of holes ($P_h$<0.5), the integrand of Eq.(2) reduces slower than expected for short chains affecting significantly the determined viscosity. Thus, $\eta_0$ for long chains is smaller than expected and for short chains larger than expected. Since eventually for large $N$, viscosities converge to the values originally predicted in the reptation theory, $\beta$ tends to be larger for chains with a great number of holes ($P_h$=0.9) and smaller for chains with a small number of holes ($P_h$=0.1).

Similar results to those reported above can be obtained analyzing the repton model. This model consists also of $N$ beads and jumps are similar to those in the necklace model [16]. In the repton model the jumping probabilities are $p_b=p_c=1/z$ and $p_a=(z-1)/z$ ($z$ reflects the dimension of the problem). Hence, the end beads always have an effective friction smaller than that of the middle beads ($p_a+p_b > p_c$) and then a chain cannot have the original reptation scaling for diffusivity. In the repton model the distance between beads can be 0 or 1 and, consequently, chains are shorter than in the necklace model. Consistently, the effects of jump distance, lack of proportionality between length and mass of the chains, and fluctuations are expected to be stronger. Indeed, Monte Carlo results show the expected trends due to these effects and they can be explained in a similar way to those observed in the necklace model.

In the range $20>N>100$, an effective value of $\beta$ can be determined to be 3.31 for $z=6$ (the exact value depends on the specific range used in the fitting). As in the necklace model, we feel compelled to attribute these results to fluctuations. Consistently, a rigid chain, moving as dictated by the center of mass in the repton model, shows a viscosity always larger than that for the flexible chain and in the range $20<N<100$, $\beta$ is almost 3. As for the necklace model, these findings can be shown to be consequence of the fluctuations details. We have checked the effect of using $z<2$ and the viscosity becomes larger, especially for small $N$, so that $\beta$ tends to be smaller than 3, analogously to what happens with $P_h=0.1$ in the necklace model.

## Conclusions

In this work we have analyzed in detail the effects on the viscosity of discretizing the reptation model as originally proposed using the necklace model. The inherent simplicity of the model allows us to quantify the consequences of discretization and fluctuations showing not expected results. In order to do this, we proposed some simplified models. We first determined the consequences of jump distance and temporal discretizations. Distance discretization lifts up viscosity and is noticeable for short chains; temporal discretization does not alter results substantially. In the necklace model jumps are discrete, and then discretization effects are unavoidable.

Then, we quantified the effects due to the lack of proportionality between the length and the mass of the chains being inherent to the studied model and found it significant. We also studied the effects of the jumping chain not being a random walk and found that only affects the viscosity non-negligibly of very short chains.

We finally addressed the effect of length fluctuations. It was shown that chain length fluctuations can affect viscosity in an unexpected way. It is regularly argue that due to length fluctuations, viscosity becomes smaller. Surprisingly, the comparison with the resulting viscosity for a rigid-length chain, with the length corresponding to the necklace model, indicates that fluctuations can increase or decrease the viscosity exponent with a variety of effects on the viscosity exponent depending on the used parameters. The found results show that the manner in which fluctuations behave can be crucial in this type of models. For small values of *N*, say smaller than 10, discretization strongly affects the resulting viscosity. For larger values of *N*, the absolute values of viscosity become relatively close to those expected but the slope $\beta$ continues depending on the model details even for *N*>100. The present analysis can be easily extended to similar models.

This work was partially supported by the Consejo Nacional de Investigaciones Científicas y Técnicas, Argentina (CONICET) and the Comisión de Investigaciones Científicas de la Provincia de Buenos Aires, Argentina (CIC).

# Figure Captions

**Fig. 1**   Chain of five beads showing possible hopping in the necklace model. Grey particles can jump as indicated with the arrows with probabilities $p_a$, $p_b$, and $p_c$ according to the configuration. White particles cannot move because jumps are only possible towards a neighboring hole and only one hole is allowed between particles.

**Fig. 2**   Viscosity in the rigid chain model as a function of the chain length $N$. The straight line is proportional to $N^3$. The inset shows $\eta_0/N^3$ as a function of $N$. For $N>10$ the curve becomes horizontal indicating that $\beta \to 3$ in accord with the original reptation model.

**Fig. 3**   Viscosity in the rigid chain model for different jump distance $\Delta x$ keeping the diffusivity constant. The jump distance discretization is responsible for the anomalous viscosity observed for small values of $N$.

**Fig. 4**   Viscosity for the equivalent length model using chain lengths corresponding to the necklace model. Lines correspond to the expected effects reflected in Eqs. (6, 7). Note that the small lack of proportionality between $L$ and $N$ strongly reflects on the viscosity for $N<10$.

**Fig. 5**   Equivalent length model using lengths corresponding to the necklace model with $P_h=0.9$. As the jump distance discretization effect is suppressed, Monte Carlo results converge to those corresponding to Eq. (8) (line).

**Fig. 6**   Viscosity as a function of the number of beads $N$ plotted as $\eta_0/N^3$ in a log-log plot for the necklace (full symbols) and hybrid (empty symbols) models. In the hybrid model, chain lengths do not fluctuate and their centers of mass move as dictated by the centers of mass of chains that follow the necklace model dynamics. In the range $20<N<100$, fluctuations can make the viscosity exponent larger ($P_h=0.9$) or smaller ($P_h=0.1$).

**Figure(1)**

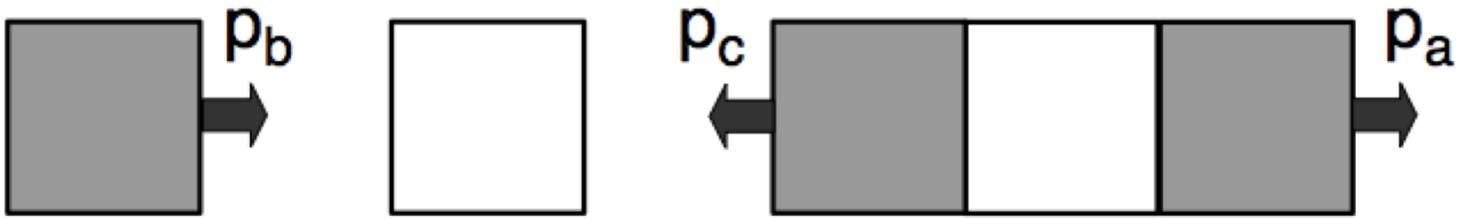

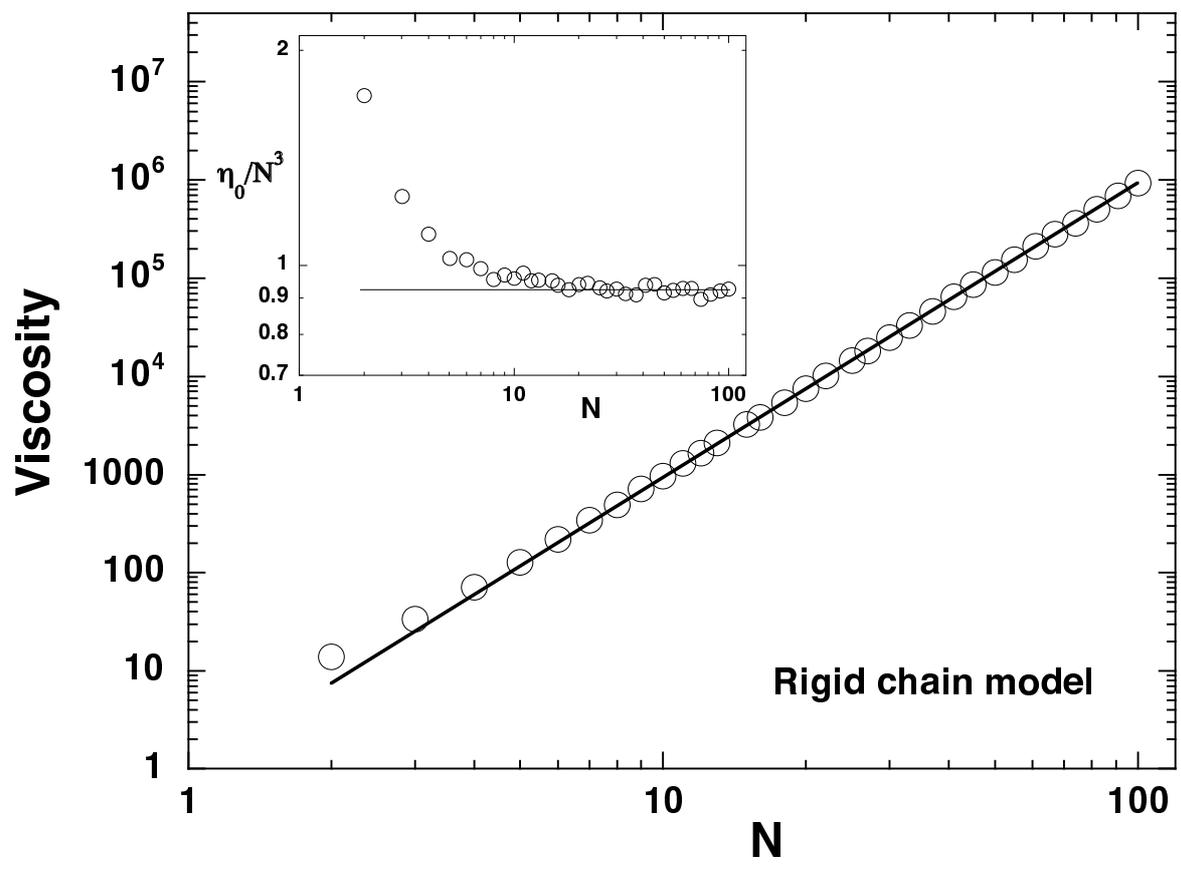

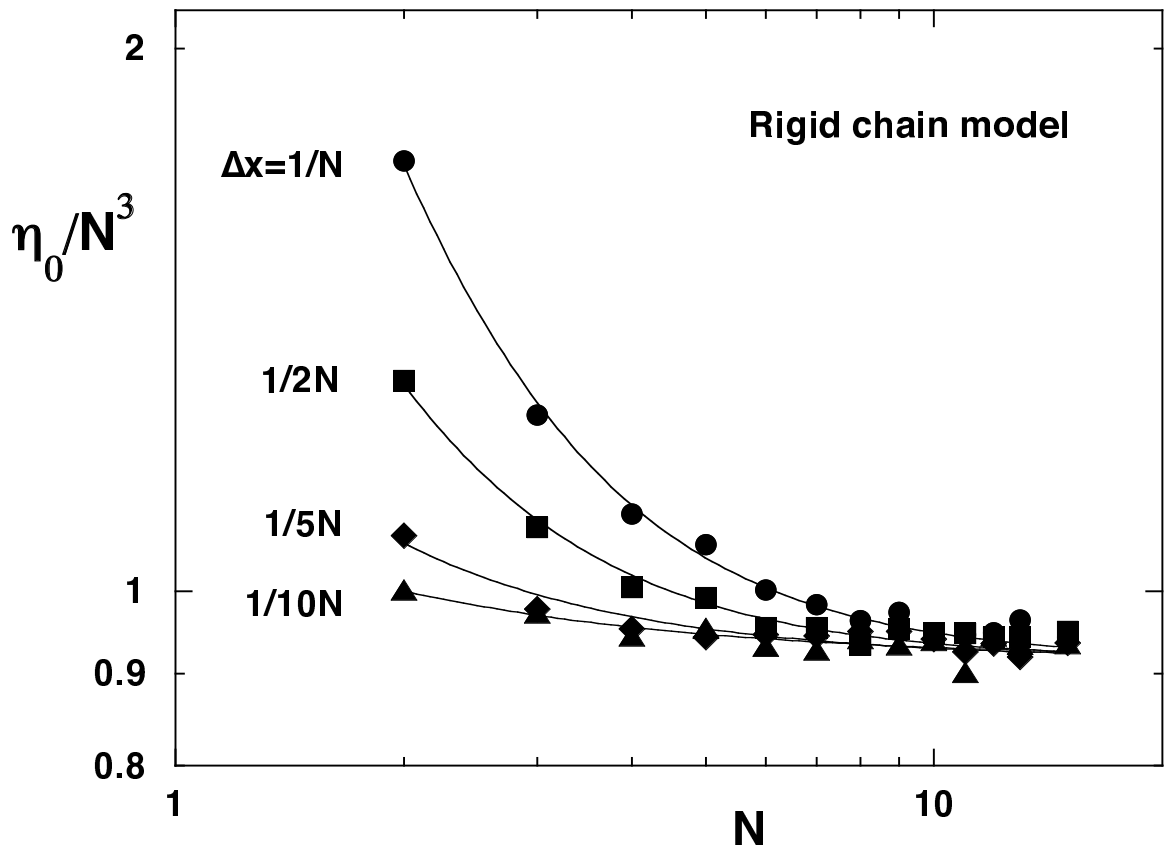

**Figure(4)**

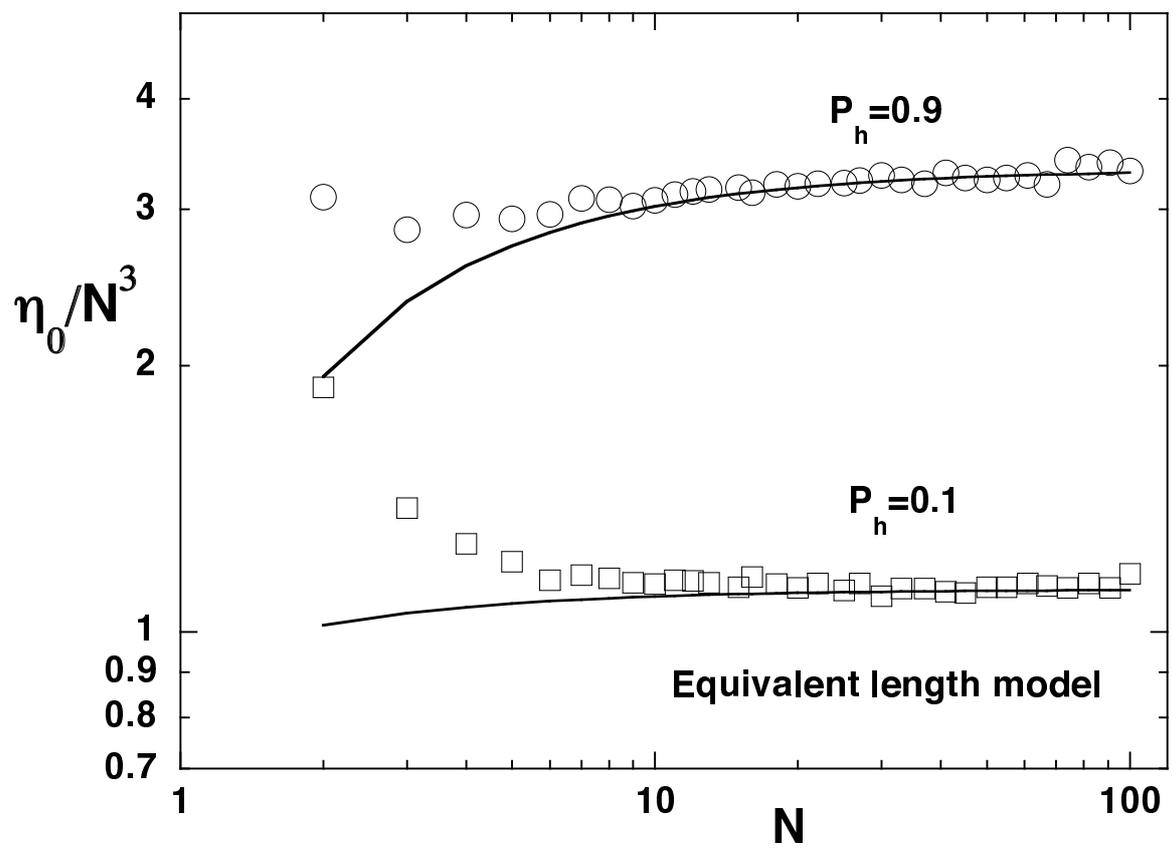

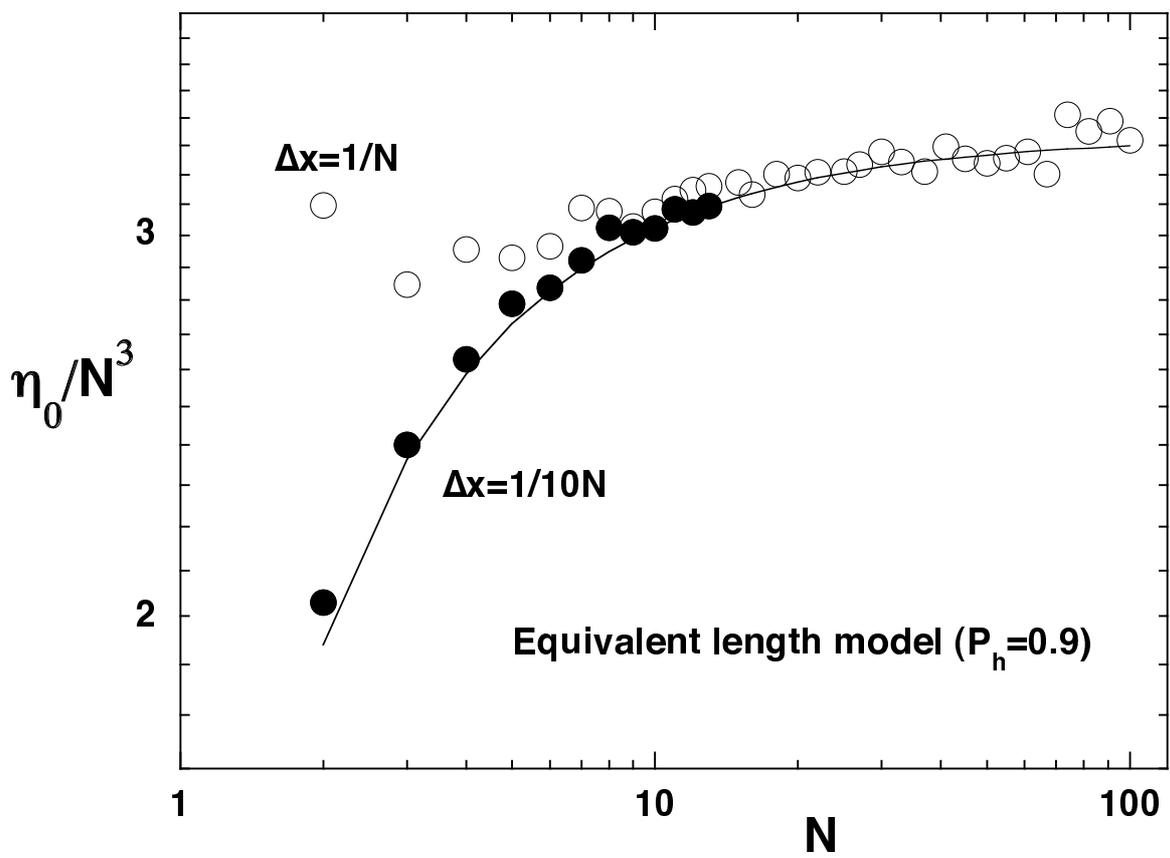

Figure(6)

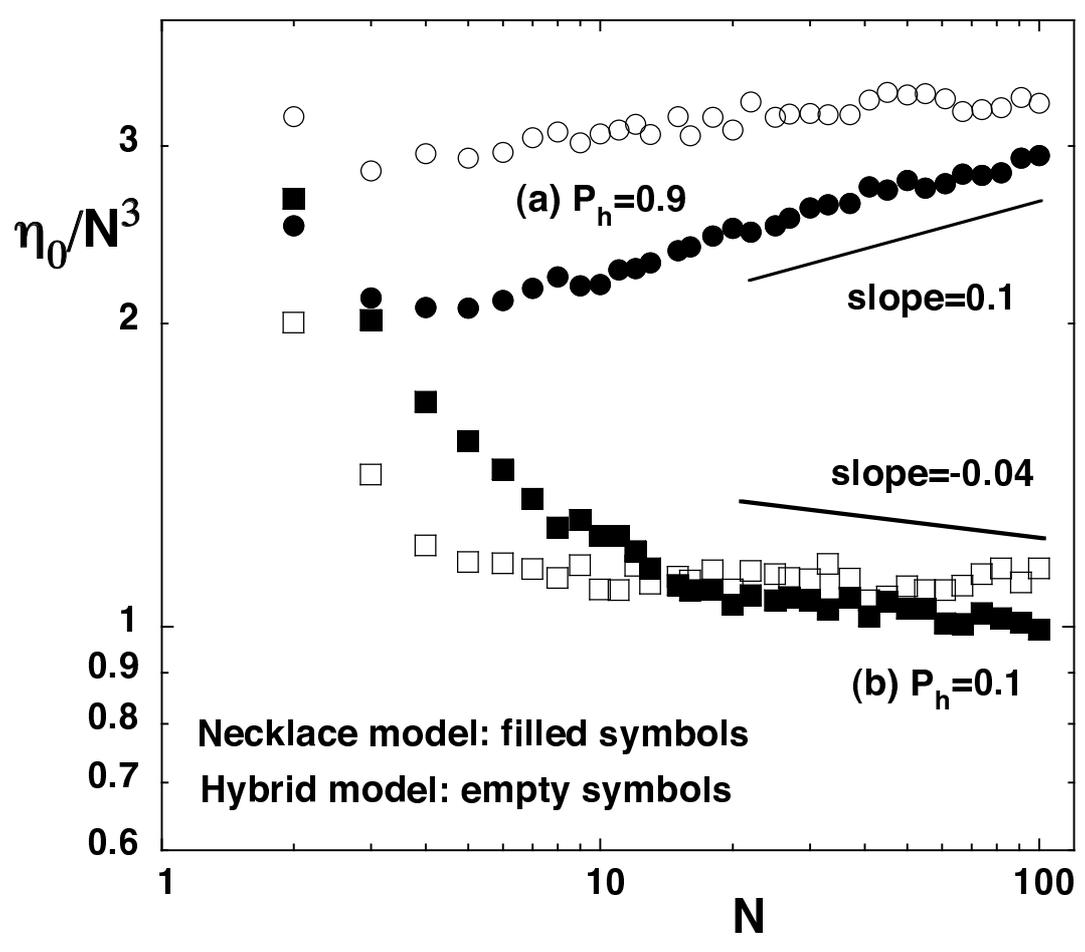